
\input phyzzx

\overfullrule0pt
\def\calo{{\cal O}}
\def\call{{\cal L}}

\def\ie{{\it i.e.}}
\def\dots{\cdotp\cdotp\cdotp}

\def\UPENN{\centerline{\it Department of Physics}\centerline{\it University
of Pennsylvania}\centerline{\it Philadelphia, PA 19104, USA}}
\nopubblock

\titlepage

\line{\hfil UPR--0603T}
\line{\hfil February 1994}
\vfill
\title{\bf Top Radiative Corrections
 in Non-minimal Standard Models}

\author{Alex Pomarol}
\UPENN
\vfill

\centerline{\bf Abstract}

We derive the one-loop effective action induced by a heavy top
in models with an extended Higgs sector. We use the effective action
to analyze the top corrections to the $\rho$ parameter and to
the Higgs-gauge boson couplings. We show that in models with
$\rho\not=1$ at tree-level, one does not lose generally the bound
on $m_t$ from the $\rho$ parameter.

\vskip .1in
\vfill

\endpage

\noindent{\bf 1. Introduction}
\REF\lan{For a recent review, see for example,
 P. Langacker and M. Luo, Phys. Rev. {\bf D44}
(1991) 817.}
\REF\lyn{B.W. Lynn and E. Nardi, Nucl. Phys. {\bf B381} (1992) 467.}
\REF\pas{G. Passarino, Nucl. Phys. {\bf B361} (1991) 351.}

Recent precision measurements at LEP allow us to strongly
 constrain the top mass in the standard model (SM)\refmark\lan.
These constraints are obtained by analyzing the  radiative
 corrections induced by the top quark to
  measurable quantities.
{}From the $\rho$ parameter, $\rho=m^2_W/(m^2_Z\cos^2\theta_W$),
we get the strongest constraint on $m_t$, since the
top radiative corrections to $\rho$
 grow quadratically with the top mass.

In the minimal SM in which the Higgs sector consists
of one Higgs doublet, the value of $\rho$ at tree level, $\rho_{_{tree}}$,
is equal to unity
so radiative corrections must be finite.
Nevertheless, when an extended Higgs sector is considered
(non-minimal SMs), one can have
$\rho_{_{tree}}\not=1$.
Since the experimental value of $\rho$ is very close to unity,
 one expects that, in such non-minimal SMs,
a simultaneous expansion in $(\rho_{_{tree}}-1)$
and $g^2m^2_t/m^2_W$ can be carried out such that the top corrections
 to $\rho$ are the same as that in the SM.
It has been recently claimed\refmark\lyn, however, that such an expansion is
meaningless, \ie, the limit $\rho_{_{tree}}\rightarrow 1$ is not continuous.
It has been argued that in these models $\rho$ is a free parameter, so it
 cannot be computed,
 but must be  extracted from the experiments.
The explicit calculation of the top  corrections to $\rho$ was carried out
in ref.~[\lyn]
and it was claimed not to be  finite. It implies that
one loses the bounds on $m_t$.
\REF\ste{For some examples in the SM, see H. Steger, E. Flores and Y.-P. Yao,
Phys. Rev. Lett. {\bf 59} (1987) 385; G.-L. Lin, H. Steger and Y.-P. Yao, Phys.
Rev. {\bf D44} (1991) 2139; O. Cheyette and M.K. Gaillard,
Phys. Lett. {\bf B197} (1987) 205; M. B. Einhorn and J. Wudka, Phys.
Rev. {\bf D39} (1989) 2758; H. Georgi, Nucl. Phys. {\bf B363} (1991) 301;
F. Feruglio and A. Masiero, Nucl. Phys. {\bf B387} (1992) 523.}
\REF\ell{Y. Okada, M. Yamaguchi and T. Yanagida, Prog. Theor.
Phys. {\bf 85} (1991) 1; J. Ellis, G. Ridolfi and F. Zwigner, Phys. Lett.
{\bf B257} (1991) 82.}

In this  paper we show using two different methods that
in non-minimal standard models the radiative the corrections to $\rho$
are finite and meaningful, even for large values of $\rho_{_{tree}}-1$.
In the particular model considered in ref.~[\lyn], we find that the bound
on $m_t$ is as strong as that in the SM.
This has also been stressed in refs.~[\lan,\pas].
In section 2,
we compute the top corrections to $\rho$ following the
effective action approach\refmark\ste.
 Such an approach is  suited to computing
radiative corrections to relations that depend on the vacuum expectation values
(VEVs)
of the scalar fields\foot{See
 ref.~[\ell], for an example in which the effective
potential is used to compute the top radiative corrections to
the Higgs mass in the minimal supersymmetric model.}.
Neither tadpole diagrams nor counterterms for the VEVs
of the scalars need
to be considered, since
the one-loop effective action is computed in the symmetric phase
-- before the electroweak symmetry breaking (ESB).
Furthermore, the effective action approach allows one to
relate the top corrections
of different low-energy processes.
In section 3, we reinforce our statement by computing
the top corrections to $\rho$
following the usual counterterm approach.
\vskip 3cm
\noindent{\bf 2. Effective action approach}
\REF\col{S. Coleman and E. Weinberg, Phys. Rev. {\bf D7} (1973) 1888.}

The effective action, $\Gamma(\phi)$, is defined as
the generator of the one particle irreducible (1PI) $n$-point Green's
 functions, $\Gamma^{(n)}$,
$$ \Gamma(\phi)=\sum_n{1\over n!}\int d^4x_1\dots d^4x_n
\Gamma^{(n)}(x_1,\dots, x_n)
\phi(x_1)\dots\phi(x_n)\, .\eqn\action$$
An alternative expansion of the effective action is in powers of momentum
about the point where all external momenta vanish,
$$\Gamma(\phi)=\int d^4x [-V(\phi)+\coeff{Z(\phi)}{2}\partial_\mu\phi
\partial^\mu\phi+\,
\dots]\, ,\eqn\momentum$$
where $V(\phi)$ is the so-called effective potential\refmark\col
$$V(\phi)=-\sum_n{1\over n!}\Gamma^{(n)}(p_i=0)\, \phi^n\, .\eqn\potential$$
\REF\wei{S. Weinberg, Phys. Rev. {\bf D19} (1979) 1277; L. Susskind, Phys. Rev.
{\bf D20} (1979) 2619; P. Sikivie et al., Nucl. Phys. {\bf B173} (1980) 189.}

Let us now consider the model of ref.~[\lyn].
The Higgs sector consists of a Higgs doublet with $Y=1$ and
a real Higgs triplet with $Y=0$,
$$\Phi=
\left(\matrix{\phi^+\cr \coeff{1}{\sqrt{2}}(\phi+iG^0)}\right)\ \ \ \ {\rm and}
\ \ \  \Sigma=\left(\matrix{\Sigma_+\cr \Sigma_0\cr \Sigma_-}\right)
\, ,\eqn\vev$$
respectively. Our phase convention is such that $\Sigma_-\equiv -(\Sigma_+)^*$.
We want to analyze  the one-loop effects of a heavy
top quark.
Since our model is $SU(2)_L\times U(1)_Y$ invariant, the one-loop
effective action before the ESB
is given by (following an expansion as in eq.~\momentum)
$$\eqalign{\Gamma=&
\int d^4x\left\{-V(\Phi,\Sigma)+[1+A(\Phi^{\dag}\Phi)]
(D_\mu\Phi)^{\dag} (D^\mu\Phi)+
\coeff{1}{2}(D_\mu\Sigma)^{\dag}(D^\mu\Sigma)\right.\cr
 +&\left. B(\Phi^{\dag}\Phi)
(\Phi^{\dag} D_\mu\Phi)
((D^\mu\Phi)^{\dag}\Phi)+\coeff{1}{2}C(\Phi^{\dag}\Phi)[
(\Phi^{\dag} D_\mu\Phi)(\Phi^{\dag} D^\mu\Phi)+h.c.]+\,\dots
\right\}\, ,}\eqn\effective$$
where we have only kept terms with a maximum of two covariant
 derivatives, which are the only terms relevant
to our analysis.
Note that the operators $A(\Phi^{\dag}\Phi)$,
 $B(\Phi^{\dag}\Phi)$ and $C(\Phi^{\dag}\Phi)$ that arise at one-loop level
 only depend on $\Phi$ because $\Sigma$ does not
couple to the quarks.
When the neutral Higgs develop
VEVs,
$\VEV{\phi}\equiv v$ and
$\VEV{\Sigma_0}\equiv v_3$,
the  operators in eq.~\effective\ induce mass terms for
 the gauge bosons. The last three terms in eq.~\effective\
 contribute  differently to
 the $W$ and $Z$ masses, \ie,
they break the custodial $SU(2)$ symmetry\refmark\wei.
The Higgs triplet kinetic term only contributes to the $W$ mass,
while $B(\Phi^{\dag}\Phi)(\Phi^{\dag} D_\mu\Phi)((D^\mu\Phi)^{\dag}\Phi)$ and
$ C(\Phi^{\dag}\Phi)(\Phi^{\dag} D_\mu\Phi)(\Phi^{\dag} D^\mu\Phi)$
 contribute only to the $Z$ mass\refmark\ste.
Notice that these two terms
are finite since they
 correspond to operators of dimension higher than 4.
The first two terms in eq.~\effective, however, are not finite.
 The effective potential  $V(\Phi,\Sigma)$ can be
renormalized following ref.~[\col]. The kinetic term for the Higgs doublet
can be made finite by a
 field redefinition
$$\Phi\rightarrow \left.(1-A)^{1/2}\right|_{\Phi=\VEV{\Phi}}\ \Phi\, .
\eqn\norma$$
After the rescaling \norma\ and the renormalization of the
effective potential, the one-loop effective action
is finite.
\REF\gol{M. Golden, Phys. Lett. {\bf B169} (1986) 248; G. Passarino,
Phys. Lett. {\bf B231} (1989) 458; {\bf B247} (1990) 587.}
\REF\vel{D.A. Ross and M. Veltman, Phys. Lett. {\bf B95} (1975) 135.}
\REF\per{S. Peris, Mod. Phys. Lett. {\bf A6} (1991) 1505.}

As a function of the neutral Higgs and gauge bosons,
 the effective action \effective\
before the redefinition \norma\ is given by
$$\eqalign{\Gamma=\int d^4x \Bigl\{&-V(\phi,\Sigma_0)+
\coeff{Z(\phi^2)}{2}\partial_\mu\phi\partial^\mu\phi+
\coeff{1}{2}\partial_\mu\Sigma_0\partial^\mu\Sigma_0\cr
&+\Pi_W(\Sigma_0^2,\phi^2)W_\mu W^\mu+\coeff{1}{2}
\Pi_Z(\phi^2)Z_\mu Z^\mu\Bigr\}\, ,}\eqn\mass$$
where
$$\eqalign{Z(\phi^2)=&1+A(\phi^2)+{\phi^2\over 2}[B(\phi^2)+C(\phi^2)]\, ,\cr
         \Pi_W(\Sigma^2_0,\phi^2)=&g^2\Sigma^2_0+
{g^2\phi^2\over 4}[1+A(\phi^2)]\, ,\cr
         \Pi_Z(\phi^2)=&{g^2\phi^2\over 4\cos^2\theta_W}[1+A(\phi^2)]+
{g^2\phi^4\over 8\cos^2\theta_W}
 [B(\phi^2)-C(\phi^2)]\, .}\eqn\relation$$
The  $\Pi_W$ and $\Pi_Z$ Green's functions can be easily calculated. They
correspond to the 1PI Green's functions with two external $W$ or $Z$
and arbitrary number of external $\phi$.
To one top-bottom loop order, they are given by
$$\eqalign{\Pi_W=&g^2\Sigma^2_0+{g^2\phi^2\over 4}
+{g^2N_c\over 32\pi^2}\left[\sum_{i=t,b}
m^2_i\left(\Delta-\ln{m^2_i\over \mu^2}+{1\over 2}\right)
-{m^2_tm^2_b\over m^2_t-m^2_b}\ln{m^2_t\over m^2_b}\right]\, ,\cr
\Pi_Z=&{g^2\phi^2\over 4\cos^2\theta_W}+{g^2N_c\over
32\pi^2\cos^2\theta_W}\left[\sum_{i=t,b}
m^2_i\left(\Delta-\ln{m^2_i\over \mu^2}\right)\right]\, ,}\eqn\pis
$$
where
$$m_{t,b}={h_{t,b}\over \sqrt{2}}\phi\, ,\eqn\mtb$$
$N_c$ is the colour number ($N_c=3$ for quarks),
$\mu$ is the renormalization constant
and  $\Delta=\ln4\pi-\gamma+1/\epsilon$, where $\gamma$ is the Euler
constant and $\epsilon=(4-n)/2$ with $n$ being the space-time dimension.
{}From eqs.~\relation\ and \pis, we can extract $A(\phi^2)$ and
 $[B(\phi^2)-C(\phi^2)]$\foot{To obtain the explicit form of $B(\phi^2)$
 and $C(\phi^2)$ we need to calculate  $Z(\phi^2)$. Nevertheless, only
the difference $[B(\phi^2)-C(\phi^2)]$ is relevant to our analysis.}.
We now rescale
 the neutral Higgs doublet as in eq.~\norma, and we obtain
$$\eqalign{\Pi_W=&g^2\Sigma^2_0+{g^2\phi^2\over 4}\, ,\cr
\Pi_Z=&{g^2\phi^2\over 4\cos^2\theta_W}-{m^2_W\over
\cos^2\theta_W}\Delta\rho_{tb}\, ,}\eqn\pisred$$
where
$$\Delta\rho_{tb}\equiv -{g^2\phi^4\over 8m^2_W}[B(\phi^2)-C(\phi^2)]
={g^2N_c\over 32\pi^2m^2_W}
\left[{1\over 2}(m^2_t+m^2_b)-{m^2_tm^2_b\over m^2_t-m^2_b}
\ln{m^2_t\over m^2_b}\right]\, .\eqn\deltatb$$
Thus, the $\rho$ parameter is given by
$$\rho=\left.{\Pi_W(\Sigma^2_0,\phi^2)\over \Pi_Z(\phi^2)\cos^2\theta_W}
\right|_{\rm VEV}=\rho_0 (1+\rho_0\Delta\rho_{tb})\, ,\eqn\rhof$$
with
$$\rho_0=(1+{4v_3^2\over v^2})\, .\eqn\rhozero$$
In eq.~\rhozero\ $v$ and $v_3$ are renormalized quantities (the values
of $\phi$ and $\Sigma_0$ that
minimize the renormalized effective potential), so
 the radiative corrections to the $\rho$
parameter are   finite and meaningful.
In our  particular non-minimal SM,
we have, from the experimental value of the $\rho$ parameter\foot{
Since the leading top contributions [eq.~\pis] do not depend on the energy
 scale, $\rho$ can be extracted from experiments at the $m_Z$ scale or from
low-energy experiments such as neutrino scattering. Our value of $\rho$ is
taken from ref.~[\lan].}, $\rho=1.005\pm 0.0024$,
 a  stronger upper bound
on $m_t$ than that in the SM since both contributions (from the Higgs triplet
and the top) are positive. In the limit $v_3\rightarrow 0$, we
get the SM prediction.
We can write eq.~\rhozero\ as a function of only $v_3$ using the relation
$$\left.{G_F\over \sqrt{2}}={g^2\over 8\Pi_W(\Sigma^2_0,\phi^2)}
\right|_{\rm VEV}={1\over 2[v^2+4v^2_3]}\, ,\eqn\gfermi$$
where $G_F$ is the Fermi constant measured from the $\mu$-decay.
Explicitly,
$$\rho_0=1+4\sqrt{2}G_Fv^2_3\, ,\eqn\rhozerotwo$$
that implies
$$v_3<7.8\ {\rm GeV}\, .\eqn\boundv$$
In models with a non-minimal Higgs sector, large radiative corrections
can also be induced by Higgs bosons\refmark\gol. In the model
\vev, however, we have noted
 that, neglecting terms of $\calo(v_3/v)\sim 3\cdot 10^{-2}$,
the Higgs sector has an approximate
global $SU(2)$ custodial symmetry under which $\Sigma$ transforms as a triplet.
It follows that Higgs corrections to $\rho$ are
 very small and the bound \boundv\ holds.
It is important to note that
eq.~\rhof\ is a result valid for any non-minimal SM.
In a general case,
$$\rho_0={\sum_i(T^2_i-T^2_{3i}+T_i)|\VEV{\phi_i}|^2\over
\sum_i2T^2_{3i}|\VEV{\phi_i}|^2}\, ,\eqn\rhog$$
where $T_i$ and $T_{3i}$ are the total and third component of the
weak isospin of $\phi_i$.
As is well known\refmark\vel,
 for a  SM with an additional complex Higgs triplet
with $Y=2$, $\chi$,
 we have
 $\rho_0=1-4\sqrt{2}G_F\VEV{\chi}^2$ for small values of $\VEV{\chi}$.
Then, a partial cancellation can take place
 between the terms $4\sqrt{2}G_F\VEV{\chi}^2$ and
 $\Delta\rho_{tb}$ so that a
larger $m_t$ is allowed in this model.
For a very heavy top, however,
a non-perturbative calculation of $\rho$ is necessary.
Such a calculation was carried out in ref.~[\per] using
a $1/N_c$ expansion.

The Higgs effective potential, once renormalized,
 depends on $m_t$. Then, if  $v_3$ is obtained
from the
minimization conditions of the effective potential, $v_3$ will depend on $m_t$.
One would expect
$$v_3^2(m_t)=v_3^2(m_t=0)+\Delta\, ,\eqn\merda$$
where $\Delta$ is of $ \calo(g^2m^2_t)$ or even of
 $\calo(g^2m^4_t/m^2_W)$, \ie,
the smallness of $v_3$
is not stable under radiative corrections of a heavy top.
It is easy to see, however, that this cannot be the case.
Consider the most general  Higgs potential\refmark\vel
$$V(\Sigma_0,\phi)=a_1\Sigma^2_0+a_2\Sigma^4_0+
a_3\Sigma^2_0\phi^2+a_4\Sigma_0\phi^2+V(\phi)\, .\eqn\potentialn$$
{}From the minimization condition of eq.~\potentialn, we have
$$v_3(m_t=0)\simeq {-a_4v^2\over 2[a_1+a_3v^2]}\, ,\eqn\vthree$$
where $v_3$ has been assumed to be small.
Because $\Sigma$ does not couple to the top, there is no vertex correction
to  $a_i$. The only correction arises from the redefinition
of the Higgs doublet \norma. Thus,
$$V^{1-loop}(\Sigma_0,\phi)=a_1\Sigma^2_0+a_2\Sigma^4_0+
a_3(1+\Delta)\Sigma^2_0\phi^2+a_4(1+\Delta)\Sigma_0\phi^2
+V(\phi)\, ,\eqn\potentialnn$$
with $\Delta\sim \calo(g^2m^2_t)$. The explicit form of $\Delta$ depends on
how we renormalize the effective potential, \ie,
the definitions of the renormalized $a_3$ and $a_4$.
{}From eqs.~\vthree\ and \potentialnn, we have
$$v_3^2(m_t)=v_3^2(m_t=0)[1+\coeff{a_1}{a_1+a_3v^2}\Delta]\, .\eqn\dkjldf$$
Therefore, $v_3$ has a weak dependence on the top mass and on the
renormalization
prescription of the
 effective potential.
\REF\ell{J. Ellis, M.K. Gaillard and D.V. Nanopoulos, Nucl. Phys. {\bf B106}
(1976) 292; A.I. Vainshtein, V.I. Zakharov and M.A. Shifman, Sov. Phys. Usp.
{\bf 23} (1980) 429.}
\REF\cha{M.S. Chanowitz, M.A. Furman, and I. Hinchliffe,
Nucl. Phys. {\bf B153} (1979) 402; S. Dawson and S. Willenbrock,
Phys. Lett. {\bf B211} (1988) 200; Z. Hioki, Phys. Lett. {\bf B224}
 (1983) 284.}
\REF\menf{A. Grifols and A. M\'endez, Phys. Rev. {\bf D22} (1980) 1725.}
\REF\men{A. M\'endez and A. Pomarol, Nucl. Phys. {\bf B349} (1991) 369;
M.C. Peyran\`ere, H.E. Haber and P. Irulegui,
Phys. Rev. {\bf D44} (1991) 191.}
\REF\hab{H.E. Haber and A. Pomarol, Phys. Lett. {\bf B302} (1993) 435.}

The one-loop effective action \effective\ gives us more information
than the top-bottom corrections to the gauge boson masses.
{}From eq.~\effective\ one can also obtain the one-loop
Higgs-gauge boson couplings. In the case of a neutral Higgs, the $\phi^nWW$
($\phi^nZZ$) coupling is given by the $n$th derivative of $\Pi_W$($\Pi_Z$)
respect to $\phi$ at $\phi=v$\refmark\ell.
For example, the
one-loop $\phi ZZ$ vertex is given by
$$\Gamma_{\phi ZZ}=
\left.{\partial\Pi_Z\over \partial \phi}\right|_{\phi=v}=
{g^2v\over 2\cos^2\theta_W}
+{g^2N_c\over 16\pi^2v\cos^2\theta_W}\left[\sum_{i=t,b}
m^2_i\left(\Delta-\ln{m^2_i\over \mu^2}-1\right)\right]\, ,\eqn\tri$$
 in agreement
with the explicit one-loop calculation\refmark\cha.
In the model \vev, the $H^+WZ$ coupling can also be obtained from
 eq.~\effective. The $H^+$ is the orthogonal state
to the charged Goldstone boson, \ie,
$$H^+=-\sin\beta\ \phi^++\cos\beta\ \Sigma_+\, ,\eqn\charged$$
where  $\tan\beta=2v_3/v$.
Eqs.~\vev--\norma, \deltatb\ and \charged\ yield
$$\call_{H^+WZ}=\left[{g^2v\sin\beta\over
 2\cos\theta_W}-{gm^2_W\sin\beta\over m_Z\cos^2\theta_W}
\Delta\rho_{tb}\right]\, H^+W_\mu Z^\mu\, .\eqn\hwz$$
Note that the $H^+WZ$ vertex at tree-level (first term of eq.~\hwz)
 is very small because it is proportional to
 $\sin\beta\sim v_3/v$. Such a proportionality to $\sin\beta$
is maintained at one-loop level (second term of eq.~\hwz),
 so top corrections to $H^+WZ$ are also small.
In models with two Higgs doublets, $\Phi_\alpha
=(\phi^+_\alpha,\ \coeff{1}{\sqrt{2}}
[\phi_\alpha+iI_\alpha])^T\ \alpha=1,2$
 (such as the minimal supersymmetric model),
the $H^+WZ$ coupling is zero at tree-level\refmark\menf\ but can be induced
to one-loop order\refmark\men.
 In these models, if we now neglect $m_b$, the one-loop effective action
is given by eq.~\effective\ replacing
$\Phi$  by $\Phi_2$, the
Higgs doublet that couples to the top, and
$\Sigma$ by $\Phi_1$.
The $H^+$ is given by
$$H^+=-\sin\beta\ \phi_1^++\cos\beta\ \phi^+_2\, ,\eqn\chargedd$$
where now $\tan\beta=\VEV{\phi_2}/\VEV{\phi_1}$. From eqs.~\effective, \norma,
\deltatb\ and \chargedd\
we obtain
$$\call_{H^+WZ}=-\left.{g^2\phi^3_2\over 4\cos\theta_W}
[B(\phi_2^2)-C(\phi_2^2)]\right|_{\phi_2=\VEV{\phi_2}}\phi^+_2W_\mu Z^\mu=
{g^3N_cm^2_t{\rm cot}\beta\over 64\pi^2m_W\cos\theta_W}H^+W_\mu Z^\mu\,
 .\eqn\hwz$$
Notice that the $H^+WZ$ vertex
 arises only
from the custodial breaking terms of eq.~\effective\refmark\hab.

\noindent{\bf 3. Counterterm approach}

Let us for simplicity assume $g^\prime=0$. In this case,
the gauge sector of the SM depends on only two independent
parameters $g$ and $v$. The conditions that we choose to fix the two
counterterms $\delta g$ and $\delta v$ are:

\noindent a) We define the $Z$ mass to be the physical mass, \ie,
$m^2_Z|_{phy}\equiv m^2_Z=g^2v^2/4$. It follows that
$$\delta m_Z=\coeff{1}{2}[v^2g\delta g+g^2v\delta v]=-A_Z\, ,\eqn\renone$$
where $A_Z$ is the coefficient of $g^{\mu\nu}$ in the
vacuum polarization tensor of the $Z$ (it corresponds to
 $\Pi_Z(\phi^2=v^2)-m^2_Z$ in eq.~\pis).
\REF\sir{A. Sirlin and R. Zucchini, Nucl. Phys. {\bf B266} (1986) 389.}

\noindent b) We identify
$\coeff{G_F}{\sqrt{2}}\equiv g^2/8m^2_W$,
which implies
$$v\delta v=-{2\over g^2}A_W\, .\eqn\rentwo$$
If we now add a Higgs triplet to the SM,
 a new parameter, $v_3$, is introduced in the model. We fix $\delta v_3$
following the renormalization
prescription of the Higgs sector of ref.~[\sir], \ie, the renormalized
$v_3$ is defined to be the true VEV of $\Sigma_0$ at one-loop.
Neglecting the mixing between the Higgs doublet and the triplet, which
is of $\calo(v_3/v)$, one finds
 $\delta v_3=0$.
Thus, eqs.~\renone\ and \rentwo\ still hold, and
 the physical $W$ mass is given by
$$\eqalign{m^2_W|_{phy}=&
m^2_Z+\delta m^2_Z+g^2v_3^2+2v^2_3g\delta g+A_W\cr
=&m^2_Z\left\{\left[1+{A_W-A_Z\over m^2_Z}\right]+
{4v^2_3\over v^2}\left[1+{A_W-A_Z\over m^2_Z}\right]\right\}\, ,}\eqn\masss$$
and eq.~\rhof\ is recovered.
As was noted in section 2, a change in the renormalization
prescription of the effective potential (or a change in the input data)
implies a shift $v^2_3\rightarrow v^2_3[1+\Delta]$ with $\Delta\sim
\calo(g^2m^2_t)$ and then a negligible change in eq.~\masss.
\vskip .5cm
\centerline{\bf Acknowledgments}
I gratefully acknowledge conversations with  Mirjam Cveti\v c,
Jens Erler, Howard Haber, Paul Langacker, Jiang Liu and
Nir Polonsky.
I would like to thank Paul Langacker for a critical
 reading of the manuscript.
This work was supported  by the Texas National Laboratory Research
Commission grant
\#RGFY93-292B.

\refout
\end